\documentclass{jfm}
\usepackage{graphicx}
\usepackage{epstopdf, epsfig}

\usepackage{amsmath}
\usepackage{amssymb}

\usepackage{mmap} % fixes ligatures - searchable in pdf
\usepackage[utf8]{inputenc} %
\usepackage{lineno} % Line numbers
\usepackage{color}
\usepackage{todonotes}

\usepackage{makecell}

\usepackage{tikz}
\usepackage{color}
\newcommand{\captline}{\tikz[baseline=-0.6ex]\draw[thick] (0,0)--(0.5,0);}
\newcommand{\captdotline}{\tikz[baseline=-0.6ex]\draw[thick,dotted] (0,0)--(0.54,0);}

\newcommand{\captchainline}{\tikz[baseline=-0.6ex]\draw[thick,dash dot dot] (0,0)--(0.5,0);}

\definecolor{pred}{RGB}{210, 0, 46}
\definecolor{pblue}{RGB}{104, 153, 174}
\definecolor{porange}{RGB}{245, 143, 41}
\definecolor{pgreen}{RGB}{0, 150, 105}
\definecolor{pgrey}{RGB}{95, 95, 95}
\definecolor{pgrey2}{RGB}{153, 153, 153}

\shorttitle{Nonlinear resonances in sloshing experiments}
\shortauthor{B. B\"auerlein and K. Avila}

\title{Phase-lag predicts nonlinear response maxima in liquid-sloshing experiments}

\author{
    Bastian B\"auerlein\aff{1,2,3}
    \and Kerstin Avila\aff{1,2,3} \corresp{\email{kavila@uni-bremen.de}}
}

\affiliation{
    \aff{1}Center of Applied Space Technology and Microgravity (ZARM), University of  Bremen, Am Fallturm 2, 28359 Bremen, Germany
    \aff{2}University of Bremen, Faculty of Production Engineering, Badgasteiner Strasse 1, 28359 Bremen, Germany
    \aff{3}Leibniz Institute for Materials Engineering IWT, Badgasteiner Strasse 3, 28359 Bremen, Germany
}

\begin{document}

\maketitle

\begin{abstract}

Mass-spring models are essential for the description of sloshing resonances in engineering. By experimentally measuring the liquid's centre of mass in a horizontally oscillated rectangular tank, we show that low-amplitude sloshing obeys the Duffing equation. A bending of the response curve in analogy to a softening spring is observed, with growing hysteresis as the driving amplitude increases. At large amplitudes, complex wave patterns emerge (including wave-breaking and run up at the tank walls), competition between flow states is observed and the dynamics departs progressively from Duffing. We also provide a quantitative comparison of wave shapes and response curves to the predictions of a multimodal model based on potential flow theory \citep{Faltinsen2009} and show that it systematically overestimates the sloshing amplitudes and the hysteresis. We find that the phase-lag between the liquid's centre of mass and the forcing is the key predictor of the nonlinear response maxima. The phase-lag reflects precisely the onset of deviations from Duffing dynamics and -- most importantly -- at resonance the sloshing motion always lags the driving by $90^\circ$ (independently of the wave pattern). This confirms the theoretical  $90^\circ$-phase-lag criterion \citep{Cenedese2020}. 
\end{abstract}

% Authors should not enter keywords on the manuscript, as these must be chosen by the author during the online submission process and will then be added during the typesetting process (see http://journals.cambridge.org/data/\linebreak[3]relatedlink/jfm-\linebreak[3]keywords.pdf for the full list)

\section{Introduction}

\begin{figure}
  \centering
  \includegraphics[width=13.5cm]{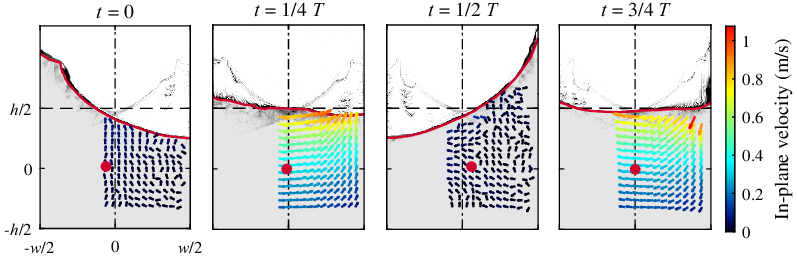}\\
  \caption{
  Sloshing liquid in a horizontally oscillated rectangular tank over one oscillation period. The tank has a width of $w=500\,\mathrm{mm}$, is filled with water to the height $h = 400\,\mathrm{mm}$ and driven with the frequency $ \omega=2\pi f$ (with $f=1.13\,\mathrm{Hz}$) 
leading to the period $T=2\pi/\omega$. Nonlinear resonances amplify periodic surface waves (marked as a red line) and produce oscillations of the liquid's centre of mass (indicated by \textcolor{pred}{$\bullet$}). Stereoscopic PIV measurements of the the in-plane velocity (displayed as arrows) show that the maximum velocities are reached when the surface elevation is lowest. The excitation frequency is close to resonance ($\Omega= \omega/\omega_1=0.917$, where $\omega_1$ is the natural frequency calculated with potential theory, see \eqref{eq:omega_n}) and the excitation amplitude corresponds to $A=x_a/w=0.64\%$, where $x_a$ is the peak amplitude of the tank displacement.}
  \label{fig:piv}
\end{figure}

Fluid resonances occur in nature and in engineering applications. One of the most prominent example is the sloshing motion of surface waves, which arise e.g.\ in partially filled tanks subjected to vibrations. As shown in figure~\ref{fig:piv}, such surface waves generate oscillations of the liquid's centre of mass. In skyscrapers this effect is used to damp vibrations after earthquakes \citep{Celebi2013}, but in rockets the sloshing fuel can threaten the mission goal, for example in the Saturn SA-I \citep{Abramson1966} and in the Falcon I Flight 2 \citep{Dreyer2009}.

In the vicinity of the primary resonance frequency, several modes of motion can compete \citep{Chester1968a, Chester1968b} and the sloshing motion can exhibit complex flow patterns \citep[e.g.\ breaking waves or swirls, see][]{Abramson1966a}. Models to predict the sloshing dynamics are essential;  \citet{Dodge2000} argued that `even with super computers, coupling the equations of motion of a flexible space vehicle to the equations of motion of a continuous liquid is too computationally demanding'. Since the early work of \citet{Moiseev1958} and \citet{Bauer1966}, mass-spring models are commonly used to predict resonances in aerospace engineering  \citep[e.g.\ in the SLOSH code of the NASA, see][]{Dodge2000}. During the critical time for the flight stability of spacecrafts, the fuel level of tanks is high, which implies that the primary resonance is typically dominated by a single mode and its higher harmonics \citep{Dodge2000, Arndt2008}. In shallow systems, on the other hand, the natural frequencies are often (nearly) commensurable, leading to a large number of excited modes and a fragmentation of the primary resonance into multiple resonance peaks.  \citet{Chester1968b} showed experimentally that several solutions with different wave morphologies coexist near the primary resonance. The number of coexisting solutions at a certain excitation frequency increases for decreasing filling level.  As the filling level tends to zero, the corresponding resonance peaks become closer, which is suggestive of a chaotic evolution \citep[see the review article of][]{Ockendon2001}. Only advanced multimodal models \citep[see e.g.][]{Faltinsen2000} are therefore able to faithfully describe the interaction of competing modes in shallow experiments \citep{Chester1968a}.

The first and most popular nonlinear mass-spring model was developed over a hundred years ago by the German engineer Georg \citet{Duffing1918}. He observed that vibrations of machines are `much calmer above resonance than at the same distance below resonance', which contradicted `the current reigning [linear] theory of these phenomena'. In order to model this dependence of the resonance frequency on the driving amplitude, Duffing added a cubic nonlinear term to the classical driven harmonic oscillator. The resulting (Duffing) equation reads 
\begin{equation} \label{eq:duffing}
 \ddot x + 2\delta \dot x + \omega_n^2 x + \epsilon x^3 = F \cos \omega t,
\end{equation}
where $x$ is the displacement of the oscillation, $\delta$ the viscous damping, $\omega_n$ the natural frequency and $\epsilon$ the  nonlinearity constant. The harmonic driving of the system is specified by the excitation amplitude $F$ and the excitation frequency $\omega$. Depending on the sign of $\epsilon$ the nonlinear resonance frequency decreases for a spring softening with growing extension 
($\epsilon<0$) and increases for a hardening spring ($\epsilon>0$). This resonance shift and the corresponding bending of the resonance curve are key features of equation \eqref{eq:duffing} and explain the original observation of Duffing  \citep[see][for a monograph on the Duffing equation]{Kovacic2011}.

Apparently unaware of the work of Duffing, \citet{Taylor1953} proposed a bended response curve resembling that of a Duffing oscillator with a softening spring to explain sloshing resonances in his experiments of a deep wavemaker tank. \citet{Tadjbakhsh1960} showed analytically and \citet{Fultz1962} experimentally that below a critical liquid depth the response curve changes from softening to hardening. This was interpreted by \citet{Chester1968a} as `a non-linear effect closely associated with the `hard spring' solution of Duffing's equation. Exploiting an inviscid theory developed by \citet{Moiseev1958}, \citet{Ockendon1973} used an asymptotic expansion of the potential flow solution in the vicinity of the resonance and showed that for small oscillations, sloshing in a two-dimensional rectangular container responds exactly as an undamped Duffing oscillator. By extending the asymptotic expansion to fifth order, \citet{Waterhouse1994} proved that fifth order terms ultimately turn the hardening into a softening. \citet{Ockendon1973} further conjectured that viscous effects emanating from the Stokes layers at the walls would lead to a linearly damped Duffing oscillator. Their conjecture has not been confirmed experimentally and the analogy between the Duffing equation and liquid sloshing has remained at a qualitative level.  

In general, the Duffing equation \eqref{eq:duffing} describes the dependence of the resonance frequency with the driving amplitude for systems with linear (laminar) damping. 
The complex phenomena occurring in large-amplitude sloshing are difficult to model because of enhanced nonlinearity and dissipation. In particular, \citet{Faltinsen2002} stated that the `improper handling of the dissipation owing to breaking waves and run-up along the vertical walls (...) limits our theory to describe the resonant sloshing of strongly dissipative character occurring in many experiments (...)'. In analogy to the Duffing equation, most models include linear viscous damping. Accurate descriptions of the dissipation and the resulting damping remain an outstanding challenge \citep{Shemer1990, Faltinsen2002, Hill2003}. For engineering applications this implies that neither the driving frequency nor the sloshing amplitude of the maximal nonlinear resonance can be predicted, if complex sloshing waves emerge. 

Accurate experiments elucidating the onset of complexity and quantifying its influence on the dissipation might provide valuable insights and stimulate the development of models. A difficulty is that experimentally the dissipation of sloshing waves cannot be measured directly. In the  literature, complex wave shapes are often shown in snapshots, but quantitative measurements related to dissipation (like the phase-lag between driving and response) are rare. In addition, the scaling of dissipation is assumed to change with the driving amplitude \citep{Faltinsen2002}, but in recent experiments response curves are seldom measured for several driving amplitudes.  
\citet{Hill2003} attributed the absence of quantitative agreement between models and sloshing experiments to the lack of well-suited and accurate experimental data. Other common problems stem from the control of the driving parameters \citep{Hill2003, Feng1997}, tanks of insufficient height \citep[leading to water impact on the tank ceiling, see e.g.][]{Faltinsen2000} and measurement procedures unable to detect hysteresis of the bended response curve \citep{Fultz1962, Arndt2008, Konopka2019}. 

In this paper, we show experimentally that sloshing at low driving amplitudes obeys the Duffing equation with linear damping, thereby confirming the conjecture of \citet{Ockendon1973}. By quantitatively measuring the flow dynamics, we link the emerging complexity at moderate amplitudes, to deviations from Duffing dynamics. We find that neither the exact surface shape, nor the frequency spectrum are useful to determine the nonlinear resonance maxima. The key indicator is the phase-lag between driving and response. We systematically investigate the role of initial conditions, characterise the sloshing amplitude with the motion of the liquid's centre of mass and directly measure the damping coefficient. The results obtained with our approach are compared to common approaches used in the literature. 

The paper is structured as follows. In the next section, we describe the experimental methods and in \S\ref{sec:character} the quantitative characterisation of the sloshing phenomena. In \S\ref{sec:duffing} and  \S\ref{sec:multimodal}, the Duffing and multimodal model of sloshing are respectively described and briefly compared to our measured data. Detailed measurements of large-amplitude sloshing are presented in \S\ref{sec:largeA} with focus on the nonlinear dynamics of the system, including multiplicity and competition of several flow states. The experimental response curves obtained for several amplitudes are presented and compared to the Duffing and multimodal model in  \S\ref{sec:effectA}. An assessment of the strengths and weakness of these two models, and of the applicability of spectral submanifold theory to sloshing, is given in  \S\ref{sec:modelling}, before the conclusions \S\ref{sec:conclusion}.

\section{Methods}\label{sec:method}
Our experiments were performed in a rectangular container subjected to harmonic horizontal excitation. As illustrated in figure~\ref{fig:piv}, the flow is quasi-two-dimensional. Sloshing waves reaching from a quasi-planar surface, up to run-up at the tank walls and wave-breaking were investigated. A distinct feature of the sloshing waves in an oscillated (or pitched) tank is their asymmetric shape leading to an oscillation of the liquid's centre of mass (shown as a red dot in figure~\ref{fig:piv}). 
Many fundamental studies consider sloshing in wavemaker tanks \citep{Taylor1953, Fultz1962, Chester1968a}. A key difference between oscillated and wavemaker tanks is that in the latter the primary resonant mode is symmetric and the liquid's centre of mass is steady in the lateral direction.

\subsection{Experimental setup}

\begin{figure}%[h]
  \centering
  \includegraphics[width=11cm]{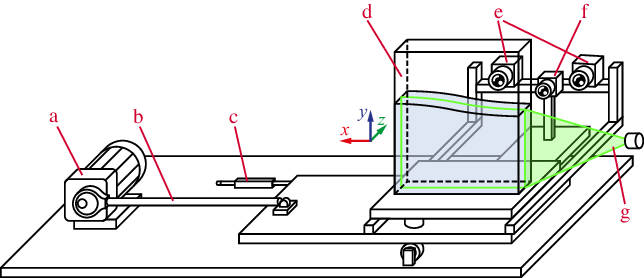}
  \caption{
  Sketch of the experiment. A motor (a) drives an eccentric disk which converts the rotary motion of the motor via a pushing rod (b) into a quasi-harmonic horizontal oscillation of the platform. A positioning sensor (c) directly records the motion of the platform on which the tank (d), two high speed cameras (e) and an USB-camera (f) are mounted. For the PIV measurements a light sheet (g) is provided by a laser passing through a cylinder lens (implemented in the stationary laser guiding arm).
  }
  \label{fig:setup} 
\end{figure}

A sketch of our experimental setup is shown in figure~\ref{fig:setup}. The tank (width $w=500\,\mathrm{mm}$, depth $l=50\,\mathrm{mm}$) is mounted on a platform and filled with water at room temperature to the height $h = 400\,\mathrm{mm}$. The tank is displaced with an excitation of
\begin{equation}
	F \cos \omega t = x_a \omega^2 \cos \omega t,
\end{equation}
where $x_a$ is the displacement peak amplitude and $\omega$ the driving frequency. The two dimensionless control parameters of the system are the excitation amplitude $A = x_a/w< 1\%$ and the excitation frequency $\Omega= \omega/\omega_1$, where $\omega_1$ is the natural frequency $\omega_n$ for a rectangular container calculated with potential theory 
\begin{equation} \label{eq:omega_n}
	\omega_n^2 = \frac{g \pi}{w} \, n \, \tanh \left[ \pi \, n \, \frac{h}{w} \right],
\end{equation} 
see \citet{Faltinsen2009}. In our experiments, the mode number is $n=1$ and $\omega_1 = 7.800\,\mathrm{s^{-1}}$. The chosen aspect ratio $h/w = 0.8$ corresponds to a deep sloshing system, i.e.\  with minor influence of the tank bottom ($\omega_n$ remains fairly constant as the aspect ratio is increased). 

The horizontal excitation of the platform is created with a three phase motor (type MDXMA1M 090-32 from Lenze), whose rotary motion is converted into a horizontal oscillation by a double eccentric and a pushing rod. The length of the rod ($1.1 \,\mathrm{m}$) is large compared to the amplitudes on the eccentric ($x_a \leq 3.180 \,\mathrm{mm}$) so that the motion of the platform can be regarded as harmonic. The driving amplitude is stepplessly adjustable by manually fixing the double eccentric. The motion of the platform is continuously monitored by a potentiometer position sensor (type 8710-100 from Burster Pr\"azisionsmesstechnik GmbH \& Co KG) with an accuracy of $\pm 5 \,\mathrm{\mu m}$. This enables precise, time-resolved measurements of the tank position and therefore also of the driving frequency, amplitude and phase. The large experimental dimensions facilitate quantitative flow measurements at low excitation frequencies $\omega=2\pi f$ (with $f\in[0.50,2.00]\,\mathrm{Hz}$) and low excitation amplitudes $x_a \in \left[0.446,3.180 \right] \mathrm{mm}$ without an influence of the tank ceiling. 
LabVIEW is used to control the motor and all the measurement equipment (except for the PIV system), which allows automatised measurements. 

Hereafter lengths are made dimensionless with the container width $w$ and time with $1/\omega_1$.

\subsection{Measuring the surface motion}
\label{cameras}

\begin{figure}
  \centering
  \includegraphics[width=13.5cm]{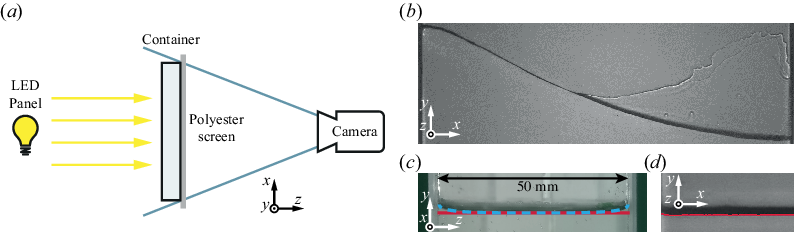}
  \caption{
  Determination of the free surface. (\textit{a}) Sketch of the visualization setup with a stationary LED-panel and a co-moving camera. (\textit{b}) In raw data images the surface appears as a thick line. A side view of the surface in (\textit{c}) allows a comparison of the detected surface level (from image processing, marked as a red line) to the real surface featuring menisci (indicted by the blue line). (\textit{d}) The image analysis eliminates the effects stemming from the menisci.
  }
  \label{fig:recording} 
\end{figure}

For presentation purposes we record the flow with our mobile phone camera ($1920\times1080\,\mathrm{Px}$ @ $60\,\mathrm{Hz}$). The camera is stationary in the laboratory frame and positioned in an angle so that the geometry and the motion of the tank are visible, too. The flow appears green in these recordings due to the laser illumination. Several movies of the sloshing are provided in the online material. 

For the quantitative analysis of the sloshing surface motion, an industrial monochrome USB-3 camera ($1600\times1200\,\mathrm{Px};\;30\,\mathrm{Hz}$) is mounted on a frame co-moving with the tank (marked as f in figure \ref{fig:setup}). The flow is illuminated from the opposite side of the tank by an LED light, which casts a distinct shadow on a semi transparent polyester film on the tank wall, see figure~\ref{fig:recording}(\textit{a}). The light is placed about $2\,\mathrm{m}$ away from the tank to provide a uniform illumination and to avoid heating the fluid. The shadow on the polyester film is monitored with a camera, which displays the water surface as a dark line, as shown in  figure~\ref{fig:recording}(\textit{b}). The apparent thickness of the line stems from the menisci at the front and back of the container wall as illustrated in figure~\ref{fig:recording}(\textit{c}). Compared to directly monitoring the free surface, this method achieves a higher contrast. The sharpness of the interface is achieved with a short exposure time ($1/200\,\mathrm{s}$) of the camera. 

The USB-3 camera is calibrated by using a checkerboard pattern calibration target and the Camera Calibrator application in MATLAB. The dark line corresponding to the water surface is identified by light intensity thresholding. To remove the effect from the menisci at the tank walls the lower bound of the dark line is defined as the relevant water surface. This is in line with the observation that at very low sloshing amplitudes the menisci seem to ``stick'' to the tank walls, whereas the free surface performs an oscillation. A locally weighted scatterplot smoothing (LOWESS) is applied to the obtained surface curve to remove outliers caused by e.g.\ small air bubbles. This method is able to detect also small surface motions and it is robust against splashes or wetting, but it is not suited to fit overturned water surfaces. Examples of the determined surface curves are shown as red lines in figure~\ref{fig:piv}. This method relies on the assumption that the flow is quasi-two-dimensional, as will be shown in \S\ref{sec:quasi2D}. 

\subsection{Measuring fluid velocities}
\label{piv}

The fluid velocities in the bulk are measured with Stereoscopic Particle Image Velocimetry (PIV). The system from \mbox{LaVision} consists of a pulsed dual-cavity $527\,\mathrm{nm}$ Nd:YAG laser $(2\times50\,\mathrm{mJ};\;1\,\mathrm{kHz})$ and two high speed cameras $(2560\times1600\,\mathrm{Px};\;1.4\,\mathrm{kHz}$, Phantom VEO 640L). The laser and a cylinder lens are stationary in the laboratory frame and provide a thin light sheet through the central plane of the tank. The cameras co-move with the tank and therefore vibrate slightly at about $12\,\mathrm{Hz}$ or faster, which is at least twice as the highest measured fluid frequencies (see {\S\ref{flow_states}}). The field of view of the cameras was varied with a maximum size of $290\times440\,\mathrm{mm}$ (as displayed in figure~\ref{fig:piv} and \ref{fig:piv_2d}(\textit{a})). Directly beneath the free surface ($\sim10\,\mathrm{mm}$) strong reflections prevented accurate PIV measurements. In most of the experiments hollow glass spheres ($9-13\,\mathrm{\mu m}$ from LaVision) were used as tracer particles. At large driving amplitudes we replaced them by silver coated hollow glass spheres ($10\,\mathrm{\mu m}$ from Dantec) to enhance the contrast. Also at large driving amplitudes highly reflective air bubbles got regularly mixed in the liquid (e.g.\ by wave-breaking) forcing us to reduce the laser intensity.  Typical sampling rates were between $200\,\mathrm{Hz}$ and $500\,\mathrm{Hz}$ and a spatial resolution of $\sim2.86\,\mathrm{mm}$ was achieved by using $64\times64\,\mathrm{Px}$ subwindows with $75\%$ overlap. 
 
\section{Quantitive characterisation of the sloshing motion}\label{sec:character}

\subsection{Assessing the quasi-two-dimensionality of the flow}\label{sec:quasi2D}

\begin{figure}
  \begin{center}
  \includegraphics[width=13.5cm]{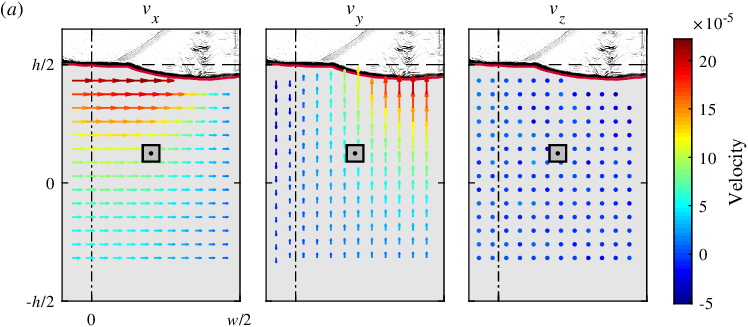}
  \includegraphics[width=6.7cm]{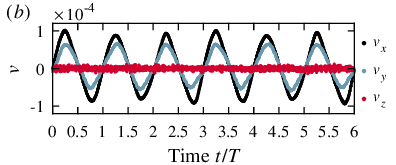}
  \includegraphics[width=6.7cm]{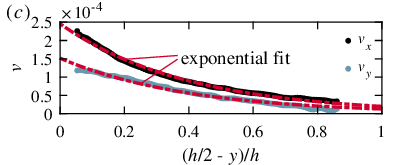}
  \end{center}
  \caption{
  Stereoscopic PIV measurements in the bulk verify the quasi-two-dimensionality of the flow. (\textit{a}) The sloshing motion is dominated by the in-plane velocities ($v_x,v_y$);  $v_z$ is negligible. (\textit{b}) Time series of the velocity components at the position marked with the black square in (\textit{a}). (\textit{c}) The in-plane velocity components decreases exponentially from the free surface to the bottom of the tank as indicated by the exponential fits. The driving parameters ($A=0.64\%$ and $\Omega=0.917$) and the flow state (period-three sloshing) are as in figure~\ref{fig:piv} (at $t=1/4T$). In (\textit{b}) and (\textit{c}) the velocities are spatially averaged over $\pm14\,\mathrm{mm}$ (size of the square in (\textit{a})).
  }
  \label{fig:piv_2d}
\end{figure}

The geometry of our tank suppresses swirling motions and to the naked eye the motion of the sloshing water surface appears quasi-two-dimensional for various driving parameters and sloshing states. In order to quantify this, we performed stereoscopic PIV measurements at the highest investigated sloshing amplitude (in which the liquid rises steeply at the tank wall). The snapshots of the velocity field shown in figure~\ref{fig:piv_2d}(\textit{a}) were recorded when the surface elevation was minimal (and the flow velocities were maximal). The in-plane velocities ($v_x$,$v_y$) reflect the sloshing motion, while the velocities in $z$-direction are negligible (and remain so throughout the cycle, see the the time series of figure~\ref{fig:piv_2d}(\textit{b})). Note that only every 8$^\text{th}$ vector obtained from the PIV measurements is plotted here for clarity. In the time series of figure~\ref{fig:piv_2d}(\textit{b}) the deviations of the $v_z$-component reflect the vibrations of the cameras itself (see {\S\ref{piv}}) and are very small, confirming that the flow is quasi-two-dimensional. For the following analysis of time series (in figure~\ref{fig:states}, \ref{fig:period3}) we applied a low-pass filter to remove these vibrations from the signal. In figure~\ref{fig:piv_2d}(\textit{c}) it is shown that the in-plane velocities decrease exponentially with the distance from the surface, as it is expected for deep water surface waves. The exponential decrease was observed at all positions.  

\subsection{Characterising the sloshing amplitude with the liquid's centre of mass} \label{sec:com_vs_surface}

\begin{figure}
  \centering
  \includegraphics[width=13.5cm]{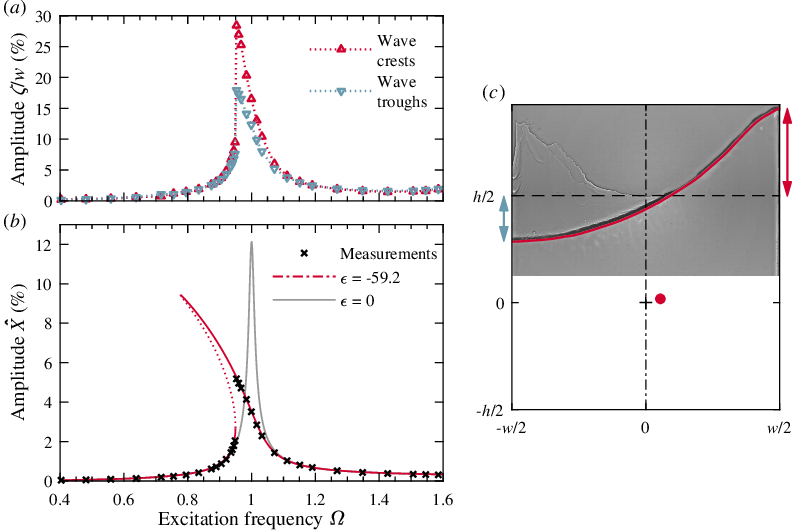}
  \caption{
  Response curves when starting the experiment from rest at $A=0.64\%$. (\textit{a}) The surface displacements of wave maxima and minima (crests and troughs) lead to different response curves of the sloshing amplitude $\zeta$. (\textit{b}) The amplitude $\hat X$ of the horizontal oscillation of the liquid's centre of mass provides a global measure of the response curve and allows a precise determination of the  sloshing phase. The solid line \textcolor{pgrey2}{\protect\captline} corresponds to a harmonic oscillator (without fitting parameters) and the red line \textcolor{pred}{\protect\captline} to a Duffing oscillator (one fit parameter, unstable solutions are denoted with dotted segments). (\textit{c}) Flow snapshot (with the free surface shown as a red line) illustrating the typical asymmetry in the surface displacement close to resonance (at $\Omega = 0.953$). This asymmetry causes the different response curves shown in (a). The liquid's centre of mass is displayed as a red dot.   
  }
  \label{fig:resonance_surface} 
\end{figure}

The most common method used in the literature to characterise the sloshing amplitude is to measure the maximum wave elevation, typically at a fixed position in the container \citep{Dodge2000,Ibrahim2005,Faltinsen2017b}. This is a direct and simple method and the data can be retrieved from wave sensors or in situ with a ruler. Using our image analysis method, we measure the surface displacement $\zeta$ with respect to the average liquid height at the sidewall. The amplitudes of the wave crests (maxima) and troughs (minima) are shown in figure~\ref{fig:resonance_surface}(\textit{a}) as a function of the excitation frequency for $A=0.64\%$. Far from the resonance (maximum response), the amplitudes of the waves and crests are similar, but close to the resonance the crests are very steep and the troughs are broad \citep{Dodge2000,Ibrahim2005}, leading to distinctly different amplitudes (as it is illustrated by the arrows in the snapshot in (\textit{c})). In many studies, it is no explicitly stated whether crests or troughs are used to characterise the sloshing amplitude, which hinders quantitative comparisons. Clearly, the surface elevation at a fixed location does not provide global information on the sloshing state, and as it will be demonstrated in the next section, this can cause problems when measuring the viscous damping coefficient.
 
In engineering applications, the motion of the centre of the liquid's mass is often the main quantity of interest (e.g.\ for the propellant tank of a satellite). The identification of the water surface combined with the quasi-two-dimensionality of the flows allows us to determine directly the centre of the liquid's mass (see the red dot in figure~\ref{fig:resonance_surface} (\textit{c})) as a function of time, $x(t)$, which we use hereafter as global measure of the sloshing amplitude. More specifically, we use the amplitude of the horizontal periodic displacement of the liquid's centre of mass, termed $\hat x$ (or $\hat X = \hat x/w$ for the dimensionless amplitude) hereafter. The corresponding response curve is shown in (\textit{b}). Being based on the analysis of the whole surface shape, this method is robust against small image evaluation errors and is well suited to quantitatively compare different flow states. 

\subsection{Determining the damping rate}
\label{sec:damping}

\begin{figure}
  \centering
  \includegraphics[width=6.7cm]{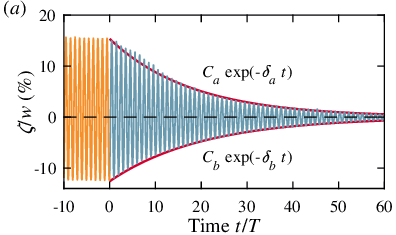}
  \includegraphics[width=6.7cm]{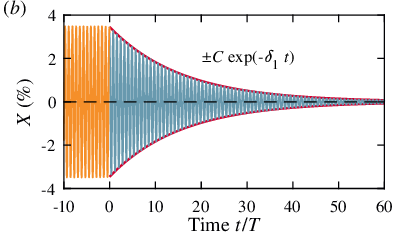}
  \caption{
  Dependence of the viscous damping rate on the observable. At $t=0$ the driving was turned off to measure the decay rate. The steady state ($t\le 0$) is a wave-breaking motion, which fades into a planar surface wave during the decay ($t>0$). (\textit{a}) Time series of the dimensionless surface displacement, $\zeta/w$, recorded at the left tank wall (where the sloshing amplitude is maximal) for $A = 0.64\%$, $\Omega = 0.999$. The asymmetry of the oscillation amplitude reflects the asymmetric wave shape (with a higher wave crest and a shallow wide trough), leading to different positive and negative displacements ($C_a = 0.153w$, $C_b = 0.126w$) and hence different decay rates ($\delta_a = 0.069\,\mathrm{s^{-1}}$, $\delta_b = 0.060\,\mathrm{s^{-1}}$).  (\textit{b}) The same time series as in (\textit{a}) but for the lateral motion of the center of mass. The displacement of the center of mass is left-right symmetric and an unambiguous damping coefficient is obtained ($C = 0.035w$, $\delta_1 = 0.064\,\mathrm{s^{-1}}$).   
   }
  \label{fig:decay_fit} 
\end{figure} 

The viscous damping rate $\delta_1$ is an important  parameter required in models of sloshing, including the Duffing oscillator and  multimodal models. In this section, it is determined experimentally and analytically for our system. In experiments, the viscous damping rate $\delta_1$ can be extracted from the system's natural response. After turning off the driving, the sloshing amplitude decays over time to zero with decay rate $\delta_1$. This is determined by fitting an exponential function to the envelope of the time series of the sloshing amplitude. In aerospace engineering it is common to use the `free surface displacement at some convenient location' \cite{Dodge2000} to obtain the damping coefficient. In figure \ref{fig:decay_fit}(\textit{a}) it is shown that this method leads to different damping coefficients for the crests and the troughs. By contrast, the motion of the liquid's centre of mass allows an unambiguous determination of the viscous damping coefficient for each time series (see figure~\ref{fig:decay_fit}(\textit{b})). Repeating such measurement for a range of excitation frequencies revealed only small deviations, $\delta_1 = [0.065\pm0.015]\,\mathrm{s^{-1}}$ (corresponding to dimensionless damping ratio $\gamma_1=\delta_1/\omega_1=8.4\cdot10^{-3}$). For very small frequencies (indicated by \textcolor{pred}{$\circ$} in figure~\ref{fig:damping}(\textit{a})) the sloshing amplitude was too small (surface wave height $< 6\,\mathrm{mm}$) to accurately fit a decaying function over the signal. As a result, these values were not considered to compute the averaged damping rate $\delta_1$ . 

\citet{Keulegan1959} derived an analytical estimate of the damping rate for liquid sloshing in rectangular containers. In his model, the damping rate is obtained from the balance of potential energy, kinetic energy and energy dissipation caused by a laminar viscous boundary-layer on the wetted tank walls and bottom. Keulegan's prediction reads \citep{Faltinsen2009},
\begin{equation} \label{eq:damping:keulegan}
	\gamma_n =\frac{\delta_n} {\omega_n}= \sqrt{\frac{\nu}{2 \, w^2 \, \omega_n}} \left[\left(\frac{w}{l}\right) \left( 1 + 2\pi \frac{n (\thalf - \frac{h}{l})}{\sinh\left[2\pi \, n \, \frac{h}{l} \right]} \right) +1 \right],
\end{equation}
where $\gamma_n$ is the dimensionless damping ratio of the natural mode $n$. In our decay experiments the first natural mode $n=1$ dominates. Plugging $\nu = 1 \cdot 10^{-6}\,\mathrm{m^2 /s}$ for water at room temperature and our experimental geometry ($h=0.4\,\mathrm{m}$, $w=0.5\,\mathrm{m}$, $l=0.05\,\mathrm{m}$) into \eqref{eq:damping:keulegan} gives $\gamma_1 = 5.57 \cdot 10^{-3}$ ($\delta_1=0.0434 \,\mathrm{s^{-1}}$), which is about $1.5$ times smaller than the experimentally determined value, $\gamma_1=8.4\cdot10^{-3}$. This difference is unsurprising, as equation \eqref{eq:damping:keulegan} only models the effect of a linear boundary-layer and does not account for damping in the fluid bulk, turbulent energy dissipation, surface tension or free-surface effects like wave-breaking \citep{Faltinsen2009}. Moreover, nonlinear effects may be relevant in the early stages of the decay, making it eventually necessary to consider also the damping of higher natural modes. As a consequence, the theory only gives a lower bound for the actual damping. Similar differences were observed in past studies \citep{Faltinsen2017a,Ikeda2012}. In our case it is noteworthy that Keulegan's prediction agrees with our measurements at low sloshing amplitude (corresponding to a low excitation frequency in figure~\ref{fig:damping}(\textit{a})), where the theory is expected to be most accurate. However, we stress that damping rates at low sloshing amplitudes could not be measured accurately. In the following, we use the experimentally determined damping ratio $\gamma_1=8.4\cdot10^{-3}$ for comparison with sloshing models.

\subsection{Determining the natural frequency}

\begin{figure}
  \centering
  \includegraphics[width=6.7cm]{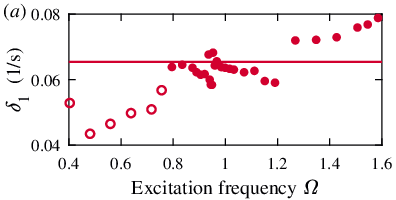}
  \includegraphics[width=6.7cm]{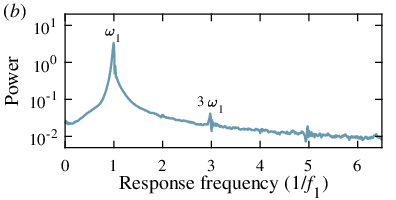}
  \caption{Measured viscous damping rate and natural frequency.
  (\textit{a}) Experimentally determined values of the viscous damping rate $\delta_1$ for different excitation frequencies and $A=0.64\%$. Their average $\delta_1= 0.065\,1/s$ is shown as a red line. (\textit{b}) Typical frequency spectrum of a decaying oscillation of the liquid's centre of mass with a peak at $\omega_1$, showing that the decaying motion oscillates with the natural frequency. The spectrum stems from the time series of the decay shown in figure \ref{fig:decay_fit}(\textit{b}). 
  }
  \label{fig:damping}
\end{figure}

The natural frequency of the sloshing can be calculated analytically with potential theory \eqref{eq:omega_n} to $\omega_1= 7.800\,\mathrm{s^{-1}}$. Experimentally we verified its value with the same measurements that were also used to obtain the viscous damping rate $\delta_1$. After stopping the forcing, the system oscillates with its natural frequency, which is seen in the frequency spectrum in figure~\ref{fig:damping}(\textit{b}). The averaged value of the measurements ($\omega_{1,\text{exp}}=[7.86\pm0.10]\,\mathrm{s^{-1}}$) agrees well with the prediction. This confirms that the measured damping rate can be attributed to the first natural mode.

\section{Modelling liquid sloshing with the Duffing equation}\label{sec:duffing}

In space engineering, the sloshing of liquid fuels is usually modelled with mass-spring models \citep[see e.g.][]{Dodge2000}. If the amplitude of the oscillations is assumed infinitesimal, the sloshing motion of the first mode is described by a classic harmonic oscillator
\begin{equation} \label{eq:harm_osc}
 m_1\left( \ddot x_1 + 2\delta_1 \dot x_1 + \omega_1^2 x_1 \right)= m_1 F \cos \omega t.
\end{equation}
where $m_1$ is the mass that participates in the sloshing motion, $x_1$ is the horizontal location of the center of the sloshing mass $m_1$, $\delta_1$ is the viscous damping rate and $m_1 F=m_1\omega^2 x_a$ is the inertial force acting in the reference frame moving with the container. It can be shown with potential theory that the sloshing mass 
\begin{equation} \label{eq:mass_fraction}
	m_n = c_n m,
\end{equation}
where $m$ is the total liquid mass and for a rectangular container 
\begin{equation} \label{eq:c_1}
	c_n= \frac{8}{\pi^3}\,\frac{\tanh\left[ \pi \left( 2n-1 \right) \dfrac{h}{w} \right]}{\left( 2n-1 \right)^3 \dfrac{h}{w}},
\end{equation}
see \citet{Ibrahim2005}. For our geometry and first eigenmode, $n=1$, we obtain $c_1 = 0.3183$, so that about $32\%$ of the total liquid mass is expected to participated in the sloshing motion.

Motivated by the conjecture of \citet{Ockendon1973} and the clear bending of the response curve shown in figure~\ref{fig:resonance_surface}($b$), we augment equation \eqref{eq:harm_osc} with a cubic displacement term resulting in the Duffing equation
\begin{equation} \label{eq:duffing_m1}
 \ddot x_1 + 2\delta_1 \dot x_1 + \omega_1^2 x_1 + \epsilon_1 x_1^3 = x_a \omega^2 \cos \omega t.
\end{equation}
In order to test the accuracy of this model equation in capturing the experimentally measured dynamics, it is crucial to note that we measure experimentally the displacement of the \emph{full} liquid's center of mass, $x$, whereas mass-spring models are concerned with the displacement of the center of mass of the sloshing liquid fraction, $x_1$. Since the static mass does not contribute to the sloshing motion, the horizontal displacement of the center of mass of the full liquid is $c_1$ times smaller than for the sloshing mass fraction. By substituting $x = c_1 x_1 $ in equation \eqref{eq:duffing_m1}, we obtain a Duffing equation for the horizontal displacement of the full liquid's center of mass
 \begin{equation} \label{eq:duffing2}
 \ddot x + 2\delta_1 \dot x + \omega_1^2 x + \epsilon x_1^3 = c_1 \,x_a \omega^2 \cos \omega t,
\end{equation}
where $\epsilon = \epsilon_1 / c_1^2$. A solution to this equation can be analytically obtained with the harmonic balance method \citep{Kovacic2011}. It reads
\begin{equation} \label{eq:hbm}
x = \hat{x} \cos \left(\omega t +\phi\right), 
\end{equation}
where the oscillation amplitude (displacement) is 
\begin{equation} \label{eq:duffing_amplitude}
	\hat{x} = \frac{c_1 \,x_a \omega^2}{\sqrt{ \left( \omega^2 - \omega_1^2 - \frac{3}{4}\,\epsilon\, \hat{x}^2 \right)^2 + \left( 2\,\delta_1\,\omega \right)^2}}
\end{equation}
and the phase difference between excitation and oscillation is 
\begin{equation} \label{eq:duffing_phase}
	\tan \phi = \frac{2\,\delta_1\,\omega}{\omega_1^2-\omega^2+\frac{3}{4}\,\epsilon\, \hat{x}^2}.
\end{equation}

The equations \eqref{eq:duffing_amplitude} and \eqref{eq:duffing_phase} are used to generate the response curves of the Duffing oscillator for our system parameters.

\subsection{Fitting the nonlinearity constant} 

The nonlinearity constant $\epsilon$ is the only parameter in equation~\eqref{eq:duffing_amplitude} which can neither be measured, nor estimated a priori with potential theory. The red line in figure~\ref{fig:resonance_surface}(\textit{b}) shows a least-square-residuals fit of equation~\eqref{eq:duffing_amplitude} to the horizontal displacement of the liquid's centre of mass, performed simultaneously to all data sets measured in this work (see \S\ref{sec:effectA}). This demonstrates that even at the large amplitude investigated here, the Duffing equation captures the measured response very well (with a single fit parameter, $\epsilon = -1.44\cdot10^{4} \,\mathrm{m^{-2}s^{-2}}$). Note that the harmonic oscillator model ($\epsilon=0$), which is entirely free of fit parameters, is able to very precisely capture the system's response far away from the resonance, see the grey line in figure~\ref{fig:resonance_surface}(\textit{b}). 

In order to ease future comparisons with numerical simulations and experiments with other dimensions, all parameters are shown hereafter in dimensionless form. With the dimensionless horizontal displacement of the liquid's centre of mass, $X=x/w$, the 
dimensionless Duffing equation reads
\begin{equation} \label{eq:duffing_dimless}
 \ddot X + 2\gamma_1 \dot X + X + \varepsilon X^3 = c_1 \, A \Omega^2 \cos \Omega t,
\end{equation}
where time  has been rescaled with the inverse of the natural frequency $\omega_1$, $\gamma = \delta_1  / \omega_1 = 8.4\cdot10^{-3}$ is the dimensionless damping coefficient and $\varepsilon=(w/\omega_1)^2\epsilon=-59.2$. A comparison between the dimensional and dimensionless parameters is shown in table~\ref{tab:duffing_params}.

\begin{table}%[h!] 
	\centering
	\begin{tabular*}{13.5cm}{l@{\extracolsep{\fill}} c c} %
		%\hline
		%\hline
		Parameter & Dimensional & Dimensionless \\
		%\hline
		Viscous damping & $\delta_1=0.065\,\mathrm{s^{-1}}$ & $\gamma_1=8.4\cdot10^{-3}$\\
		Natural frequency & $\omega_1=7.800\,\mathrm{s^{-1}}$ & $1$\\ 
		Nonlinearity constant & $\epsilon=-1.44\cdot10^{4} \,\mathrm{m^{-2}s^{-2}}$ & $\varepsilon=-59.2$\\
		%\hline	
	\end{tabular*}
	\caption{Experimentally determined parameters of the Duffing equation in dimensional \eqref{eq:duffing2} and  dimensionless \eqref{eq:duffing_dimless} form. For the measurement procedure, see \S\ref{sec:duffing}.}
	\label{tab:duffing_params}
\end{table}

\section{Multimodal model of liquid sloshing}\label{sec:multimodal}
\subsection{Model equations for sloshing in a rectangular tank}
\citet{Faltinsen2009} give a solution of the potential-flow equations for two-dimensional inviscid sloshing in a rectangular container using a Fourier  expansion. The elevation of the free-surface as a function of the horizontal container coordinate,  $\xi \in [-w/2, \, w/2]$, and of time, $t$, can be expressed as
\begin{equation} \label{eq:multi:surf}
	\zeta(\xi,t) = w \sum_{n=1}^{\infty} \beta_n(t) f_n(\xi), 
\end{equation}
where
\[ 
f_n(\xi)  = \cos\left(\pi n \left(\xi+ \thalf w\right)/w \right),
\]
is the dimensionless waveform of the surface for mode $n$ and $\beta_n$ describes its temporal evolution. When this expression is inserted in the potential-flow equations, these reduce to an infinite system of (coupled) nonlinear differential equations for the time-dependent coefficients $\beta_n$. By truncating this expansion, multimodal models of sloshing of desired order can be obtained. \citet{Faltinsen2009} provide equations for $\beta_n$ for the first three modes ($n=1,2,3$), which constitute the lowest order, consistent nonlinear multimodal model of sloshing in a rectangular container \citep{Moiseev1958}. 
The analytic solution to these equations is given in chapter 8.3 of \citet{Faltinsen2009} and reads
\begin{equation} \label{eq:multi:beta}
\begin{split}
	 \beta_1(t) & = \hat Y \cos(\omega t) + \tilde{n}_1 \hat Y^3 \cos(3\omega t) + \mathcal{O}(\hat Y^5), \\
	 \beta_2(t) & = \hat Y^2 \left(w_0 + h_0 \cos(2\omega t)\right) + \mathcal{O}(\hat Y^4), \\
	 \beta_3(t) & = \cos(\omega t) \left[\tilde{N}_1 \hat Y^3 - A P_3 / (1-(\omega_3 / \omega)^2) \right] + \tilde{N}_2 \hat Y^3 \cos(3\omega t) + \mathcal{O}(\hat Y^5),
\end{split}
\end{equation}
where $\hat Y$ is a dimensionless amplitude, $\omega_n$ are the natural frequencies of the three modes (see \eqref{eq:omega_n}). Hence, the time-dependent coefficients $\beta_n$ of the solution (up to third order in amplitude $\hat Y$) consist of harmonics of the driving frequency $\omega$, whose magnitude depends on the dimensionless amplitude $\hat Y$, and the following parameters
\begin{equation} \label{eq:multi:n_tilde}
\begin{split}
	\tilde{n}_1 & = - \frac{d_2 + h_0(3d_1 + 4d_3)}{2(9- (\omega_1 / \omega)^2)}, \\
	\tilde{N}_1 & = \frac{-3q_2 + q_4 + 2h_0(-q_1 - 4q_3 + 2q_5) - 4q_1 w_0}{4(1 - (\omega_3 / \omega)^2)}, \\
	\tilde{N}_2 & = - \frac{q_2 + q_4 + 2h_0(q_1  + 4q_3 + 2q_5)}{4 (9 - (\omega_3 / \omega)^2)},
\end{split}
\end{equation}
and 
\begin{equation} \label{eq:multi:w0_h0}
	w_0 = \frac{d_4 - d_5}{2 (\omega_2 / \omega)^2}, \quad 
	h_0 = \frac{d_5 + d_4}{2((\omega_2 / \omega)^2 - 4)}.
\end{equation}
The coefficients $d_i$ and $q_i$ depend on the ratio of filling level to container width, $h/w$, and are defined as,
\begin{equation} \label{eq:multi:d}
\begin{split}
    d_1 & = 2 \frac{E_0}{E_1} + E_1, \quad 
    d_2 = 2 E_0 \left( -1 + \frac{4 E_0}{E_1 E_2} \right), \\
    d_3 & = -2 \frac{E_0}{E_2} + E_1, \quad
    d_4 = -4 \frac{E_0}{E_1} + 2 E_2, \\
    d_5 & = E_2 - 2 \frac{E_0 E_2}{E_1^2} - 4 \frac{E_0}{E_1},	
\end{split}
\end{equation}
and 
\begin{equation} \label{eq:multi:q}
\begin{split}
	q_1 & = 3 E_3 - 6 \frac{E_0}{E_1}, \quad
	q_2 = -3 E_0 - 9 \frac{E_0 E_3}{E_1} + 24 \frac{E_0^2}{E_1 E_2}, \\
	q_3 & = 3 E_3 - 6 \frac{E_0}{E_2}, \quad
	q_4 = -6 E_0 - 24 \frac{E_0 E_3}{E_1} + 48 \frac{E_0^2}{E_1 E_2} + 24 \frac{E_0^2 E_3}{E_1^2 E_2}, \\
	q_5 & = 6 \left( \thalf E_3 - \frac{E_0}{E_1} - \frac{E_0 E_3}{E_1 E_2} - \frac{E_0}{E_2} \right),
\end{split}
\end{equation}
respectively, with 
\begin{equation} \label{eq:multi:E_P}
\begin{split}
	E_0 = \tfrac{1}{8} \pi^2, \quad E_n = \thalf \pi \tanh \left[\pi \, n \, \frac{h}{w} \right], \quad n \geq 1 \\
	P_n = \frac{2}{n \pi} \tanh \left[\pi \, n \, \frac{h}{w}\right] ((-1)^n - 1).
\end{split}
\end{equation}
Under the assumption of periodic (steady-state) oscillations and by further considering linear damping added to the equation for the first mode, \citet{Faltinsen2017a} obtained an analytic expression for the dimensionless amplitude parameter 
\begin{equation} \label{eq:multi:amplitude_Y}
	\hat Y = \frac{P_1\,A}{\sqrt{\left[ (\omega_1/\omega)^2 - 1 + M_1\,\hat Y^2 \right] + \gamma^2}},
\end{equation}
where 
\begin{equation} \label{eq:multi:nonlin_mu}
	M_1 = d_1 (-w_0 + \thalf h_0) - \thalf d_2 - 2 d_3 h_0. 
\end{equation} 
The phase-lag is
\begin{equation} \label{eq:multi:phase}
	\cos\phi = \frac{\hat Y \left[(\omega_1/\omega)^2 - 1 + M_1\,\hat Y^2 \right]}{P_1\,A}.
\end{equation}
Equations \eqref{eq:multi:amplitude_Y} and \eqref{eq:multi:phase} are used to generate the response curves of the multimodal model for our system parameters.

\subsection{Similarities and differences between the multimodal model and the Duffing equation}
The equations \eqref{eq:multi:amplitude_Y} and \eqref{eq:multi:phase} for the amplitude $\hat Y$ are very similar in structure to the solution of the Duffing equation for the amplitude of the liquid's centre of mass $\hat X$ (shown in this paper only in dimensional form, i.e.\  equations  \eqref{eq:duffing_amplitude} and \eqref{eq:duffing_phase} for $\hat x$). The square of the amplitude parameter $\hat Y$ is multiplied by a nonlinearity parameter $M_1$ akin to the Duffing parameter $\epsilon$. The response curves of the multimodal model are thus qualitatively similar to those of the Duffing oscillator, as will be shown later in \S\ref{sec:effectA}. For example, as the driving amplitude increases, the resonance peak bends and a hysteretic region with two stable solutions and an unstable solution develops. Depending on the sign of $M_1$ this resonance can be softening (bending towards smaller frequencies) or hardening (towards higher frequencies). More specifically, at a critical filling of $h_\text{crit}/w = 0.3368$ the coefficient becomes zero \citep{Ockendon1973}, with softening (hardening) predicted for larger (smaller) $h/w$. In our case $h/w=0.8$ and accordingly our system exhibits a softening response. A clear advantage of the multimodal model over the Duffing equation is that $M_1$ can be computed a priori, whereas $\epsilon$ must  be obtained with a fit to our experimental data. Note also that $M_1$ is a monotonically decreasing function of the driving frequency $\omega$. In the next section we compare experimentally measured surface shapes to the predictions of the mulitmodal model of \citet{Faltinsen2017a}.

\subsection{Comparison of the predicted and measured surface wave shapes} \label{sec:multi_waves}

\begin{figure} 
	\centering
	\includegraphics[width=13.5cm]{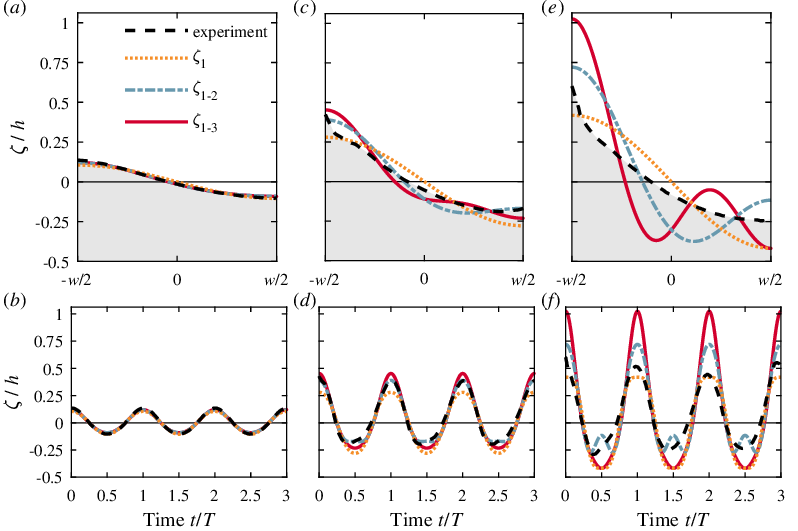}
	\caption{
	Comparison of the free-surface elevation from the multimodal model in \eqref{eq:multi:surf} to the experimentally obtained one for varying driving frequency $\Omega$ and constant amplitude $A=0.64\%$. The black dashed lines illustrate the experimentally obtained surface and the coloured lines the ones of the multimodal model for increasing number of modes as indicated in the legend in \textit{a}. The red solid line denotes e.g. the predicted surface elevation based on three modes ($N=3$).  Snapshots of the surface elevation over the container width $\xi$ are shown in the upper panel and the corresponding time series at the left wall in the panel below. For a better visual comparison the phases of the oscillations between experiment and model are optimally aligned. The driving parameters are in (\textit{a, b}) $\Omega = 0.950$, $\hat Y = 0.084$, in (\textit{c, d}) $\Omega = 0.962$, $\hat Y = 0.233$ and in (\textit{e, f}) $\Omega = 0.909$ and $\hat Y = 0.377$.
	}
	\label{fig:multi_surface_snaps}
\end{figure}

The multimodal model can be used to predict the temporal evolution of the surface shape during an oscillation period (under steady state conditions). For our geometry and selected driving parameters, the amplitude parameter $\hat Y$ is computed according to \eqref{eq:multi:amplitude_Y}, the time dependent functions $\beta_n$ are evaluated according to \eqref{eq:multi:beta} and then inserted into \eqref{eq:multi:surf}, yielding a modal prediction of the wave elevation $\zeta$. Three examples of calculated waveforms are shown in the upper row in figure~\ref{fig:multi_surface_snaps} for varying excitation frequencies. For $\Omega = 0.950$, shown in $(a)$, the amplitude parameter is $\hat Y=0.084$ and the wave shape is clearly dominated by the first mode (shown as dotted line). As the second mode is added, the prediction (dash-dotted line) becomes slightly better, but it does not improve much when the third mode is also considered (solid line).  For $\Omega = 0.962$ in $(c)$, the amplitude parameter is much larger,  $\hat Y=0.233$, and as a consequence   
the contributions of the second and third modes become more significant, which results in a steeper wave shape, a higher wave crest  and a shift of the nodal point away from the centre. These features are in qualitative agreement with our experimental measurements, however, three distinct differences are identified. In the experiments the rising of the liquid at the wall is substantially steeper, the trough is slightly more shallow and a local maximum of the surface inside the trough is not observed. For larger sloshing amplitude, $\Omega = 0.909$ and $\hat Y = 0.377$, these differences are strongly exacerbated (as shown in $(e)$), leading to very different crest maxima and qualitatively different wave shapes (with less maxima/minima in the experiments). This suggests that additional (higher order) modes or stronger damping (especially for the modes $n=2$ and $n=3$, where damping is neglected) may be necessary to accurately capture the behaviour of the system.

In the literature, the complete surface shape has rarely been quantitatively analysed experimentally, which makes it difficult to improve and validate models. Typically, the time series of the elevation at a fixed location, e.g.\ at the tank wall, is used for comparison. For completeness, the corresponding time series for the three cases above are shown in the lower panels of figure~\ref{fig:multi_surface_snaps}. For $\Omega = 0.950$ in $(b)$ and $\Omega = 0.962$ in $(d)$ an excellent agreement is obtained in both cases despite the marked differences when the whole surface is considered. Hence, our results indicate that single-point measurements might not be sufficient to reliably evaluate model predictions. We conclude that at low sloshing amplitude the wave shape of the third-order multimodal model is in excellent agreement with the experiments, whereas at larger amplitudes  the wave shape differs substantially, even if the time series at a particular points still yields a good agreement. At larger sloshing amplitudes (closer to the nonlinear resonance and displayed in $(f)$), the time series and the wave shape deviate more substantially.

\section{Nonlinear dynamics of large-amplitude sloshing}\label{sec:largeA}

\begin{figure}
  \centering
  \includegraphics[width=13.5cm]{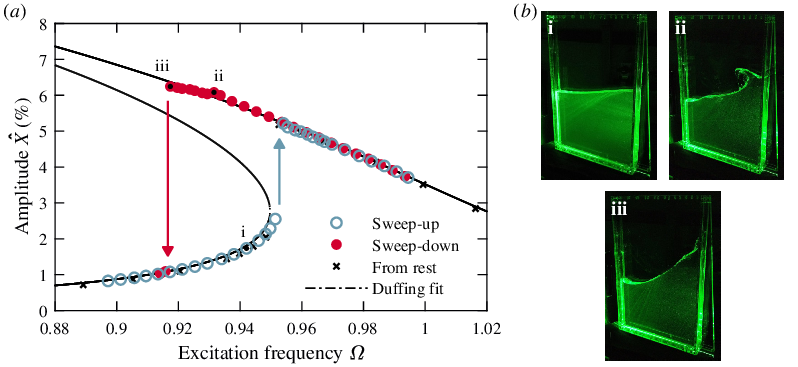}
  \caption{
  Response curve when performing frequency sweeps at $A=0.64\%$. (\textit{a}) Sloshing amplitudes obtained with quasi-static changes of the excitation frequency during a single run are marked with \textcolor{pblue}{$\circ$} for the frequency sweep-up and with \textcolor{pred}{$\bullet$} for the frequency sweep-down. The observed response curve ($\times$), when experiments were started from rest (same measurements as in figure~\ref{fig:resonance_surface}(\textit{b})) are included for comparison. Arrows indicate the (jump-up/down) transitions between the lower (low sloshing amplitude) and upper (high sloshing amplitude) branches. Hysteresis occurs in between the arrows. The line \protect\captchainline represents the fitted amplitude curve from the Duffing oscillator, with unstable solutions as dotted segments. (\textit{b}) Snapshots of different sloshing states recorded at the marked points in (\textit{a}). i: planar motion, ii: wave-breaking, iii: period-three motion. The corresponding movies of the three states (supplementary movie~1, 2, 4) are provided online.
  } 
  \label{fig:resonance_sweep}
\end{figure}

\subsection{Hysteresis}

As customary in the literature \citep{Fultz1962, Arndt2008, Konopka2019}, we started our experiments from rest and waited until transients lapsed to obtain a response curve. Subsequently, we determined the amplitude of the sloshing waves by measuring the amplitude of the horizontal periodic displacement of the liquid's centre of mass $\hat X$. As shown in figure~\ref{fig:resonance_surface}(\textit{b}) for $A=0.64\%$, this procedure yields a resonance curve with a slight asymmetry (softening), which can also be observed in the measurements of the wave height at the sidewall shown in (\textit{a}). However, starting the system form rest is insufficient to fully characterise nonlinear resonances (see e.g. \citet{strogatz2018}). In fact, a different picture emerged when the excitation frequency was changed quasi-statically in a sweep-up, sweep-down procedure (see figure~\ref{fig:resonance_sweep}(\textit{a})). This allowed it to track stable flow states through the parameter space and revealed the occurrence of a pronounced hysteresis. Compared to the measurements started from rest, the maximum sloshing amplitude occurs at smaller driving frequencies and is substantially higher. 

\subsection{Classification of flow states}
\label{flow_states}

\begin{figure}
  \centering
  \includegraphics[width=13.5cm]{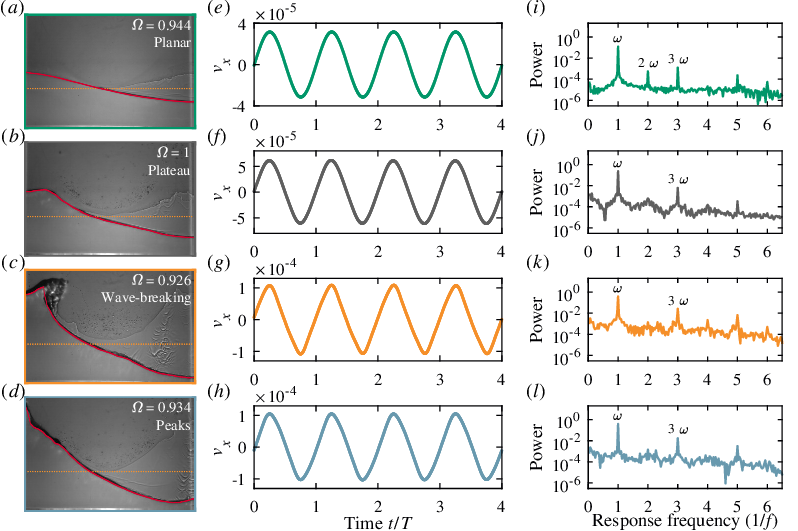}
  \caption{
  Characterisation of the sloshing states at $A=0.64\%$. (\textit{a})--(\textit{d}) Exemplary snapshots of the states. The surface is indicated with \textcolor{pred}{\protect\captline} and the surface height at rest with \textcolor{porange}{\protect\captdotline}. (\textit{e})--(\textit{h}) Time series of the horizontal velocity $v_x$ in the bulk of the flow obtained with PIV (recorded at least at $200\,\mathrm{Hz}$). The velocity is spatially averaged over $\Delta x=28\,\mathrm{mm}$ and $\Delta y=28\,\mathrm{mm}$ with the central position being at $x=0 $ and $y=0.125h$. (\textit{i})--(\textit{l}) Spectra of the time series in (\textit{e})--(\textit{h}) obtained with fast Fourier transforms. The investigated flow states are: (\textit{a},\textit{e},\textit{i}): Planar motion $\Omega=0.944$, (\textit{b},\textit{f},\textit{j}): Plateau peaks $\Omega=1$, (\textit{c},\textit{g},\textit{k}): Wave-breaking $\Omega=0.926$, (\textit{d},\textit{h},\textit{l}): Peaks $\Omega=0.934$. 
  }
  \label{fig:states} 
\end{figure}

At low excitation frequencies ($\Omega \lesssim 0.9$) the sloshing state consists of a quasi-planar wave, as exemplified in the snapshots of figure~\ref{fig:resonance_sweep}(\textit{b})-i and figure~\ref{fig:states}(\textit{a}). Its linear nature can be observed in the velocity time series in figure~\ref{fig:states}(\textit{e}), and more precisely in the frequency spectrum in (\textit{i}) dominated by the driving frequency and its higher harmonics. At low excitation frequencies, the quasi-planar wave (termed also \emph{lower branch} state hereafter) is the only stable flow state and is obtained independently of the  the initial conditions, i.e.\ start from rest or sweep-up, sweep-down. Similarly, at large frequencies ($\Omega \gtrsim 1$) there is only one stable \emph{upper branch} state, distinguished by its plateau-like wave crest near the wall, as displayed in figure~\ref{fig:states}(\textit{b}). The frequency spectrum is similar, except that the second harmonic can hardly be discerned, while the third harmonic is rather strong. For modelling approaches, these frequency spectra provide relevant information. In the multimodal model, the second harmonic, with amplitude $\mathcal{O}(\hat Y^2)$, is included in $\beta_2$ (see \eqref{eq:multi:beta}), whereas this does not appear in the Duffing equation, which features terms of amplitude $\mathcal{O}(\hat X)$ and $\mathcal{O}(\hat X^3)$ only. The weakening of the second harmonic for increasing sloshing amplitude, combined with the strengthening of the third harmonic (which appears naturally in the Duffing equation), indicates that a reduction of the dynamics to a spectral submanifold (e.g. represented by the Duffing equation) might be possible \citep{Breunung2018}.

In the following, we start from this plateau-like state (at $\Omega \sim 1$) and describe the changes in the dynamics as the excitation frequency  is decreased on the upper branch. Progressively, the surface plateau becomes broader and extends towards the centre and its leading front becomes increasingly steeper. As the frequency is further decreased, a near-vertical liquid column forms and finally it  overturns, causing wave-breaking, see the snapshots in figure \ref{fig:resonance_sweep}(\textit{b},ii) and \ref{fig:states}(\textit{c}). We define the onset of this wave-breaking state, when splashing is first observed. Its frequency spectrum is broader, reflecting the enhanced fluctuations and being suggestive of chaos. For a further reduction of the driving frequency, the wave becomes higher (without breaking) and the liquid rises steeply at the tank wall instead of overturning. We refer to this flow state shown in figure~\ref{fig:states}(\textit{d}) as peak state. Its frequency spectrum is again less broad and qualitatively similar to the plateau state at substantially lower sloshing amplitudes. 

\begin{figure}
  \centering
  \includegraphics[width=13.5cm]{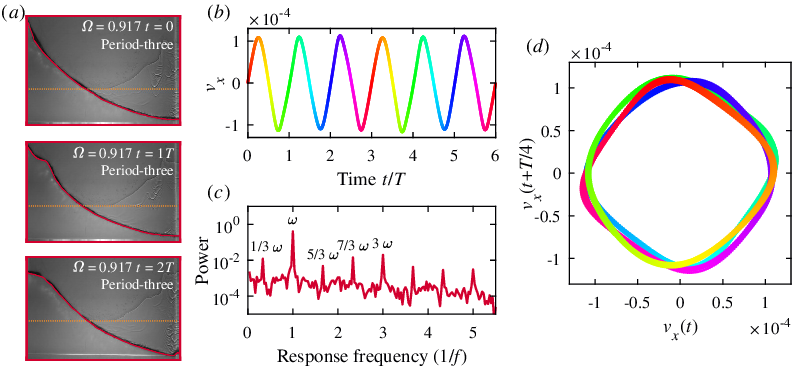}
  \caption{
  Period-three motion at  $A=0.64\%$  and $\Omega=0.917$. (\textit{a}) Snapshots of three consecutive peaks of the oscillation. (\textit{b}) Time series of the horizontal velocity $v_x$ in the bulk of the flow. The velocity is spatially averaged over $\Delta x=28\,\mathrm{mm}$ and $\Delta y=28\,\mathrm{mm}$ with the central position being at $x=0 $ and $y=0.125h$ obtained with PIV (recorded at $300\,\mathrm{Hz}$). (\textit{c}) Spectra of the time series in (\textit{b}) obtained with a Fast Fourier Transform. (\textit{d}) Phase portrait of the dynamics. The measurements correspond to 30 periods and are plotted as symbols, but appear as a line due to the low scatter in the data. A movie of the period-three motion (supplementary movie~4) is provided online.
  }
  \label{fig:period3}
\end{figure}

For a further decrease of the driving frequency, a remarkable phenomena is observed: a period-three motion where every third wave peak rises higher at the tank wall. This effect is clearly visible by naked eye (see the sequence of snapshots in figure~\ref{fig:period3}(\textit{a})) and leads to a sub-harmonic frequency $\omega/3$ and additional linear combinations in the frequency spectrum in figure~\ref{fig:period3}(\textit{c}). The resulting period-three motion is also visualised in the phase portrait in figure~\ref{fig:period3}(\textit{d}). Interestingly,  $3\omega$ and $\omega/3$ are classical secondary frequencies of the Duffing oscillator, stemming from its cubic nonlinearity \citep{Kalmar-Nagy2011}. To our knowledge these have not been observed experimentally so far, neither in sloshing fluids nor in simple systems mimicking Duffing oscillators.  
After a sufficient reduction of the driving frequency the period-three motion transitions (at the so-called jump-down transition) to the lower branch with the quasi-planar waves of low amplitude. 

\subsection{Co-existence of upper branch states in parameter space}

\begin{figure}
  \centering
  \includegraphics[width=13.5cm]{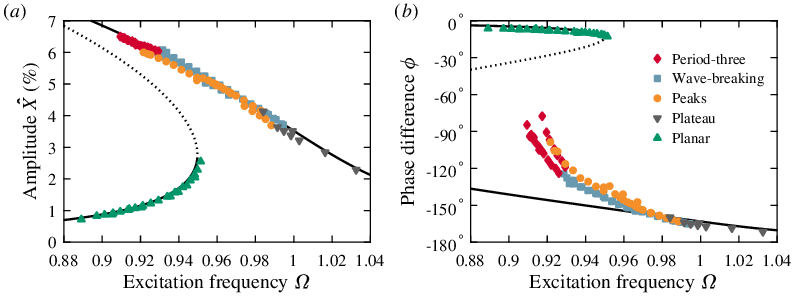}
  \caption{
  Nonlinear competition of sloshing states at $A=0.64\%$. (\textit{a}) Measured sloshing amplitudes as function of the driving frequency and (\textit{b}) corresponding phase-lag between sloshing motion and driving. The different symbols represent the observed flow state for each measurement point. The measurements in this figure were obtained with different methods (starting from rest and frequency sweeps with variable frequency increments). Differentiation between states is best seen in the phase difference. In this visualization lower and upper branches appear exchanged, because of the strongly negative phase-lag characteristic of the large amplitude sloshing branches. 
  }
  \label{fig:resonance_states} 
\end{figure}

Several repetitions of the above described measurements reveals that the flow transitions are very sensitive to experimental conditions, and that the peak state and the wave-breaking state coexist in a large span of driving frequencies, as shown in the response diagram in figure~\ref{fig:resonance_states}(\textit{a}). These flow states seem to be marginally stable, we even occasionally  observed a change from the peak state to the wave-breaking state without any change in the driving parameters. On the other hand, the plateau state at large driving frequencies ($\Omega \gtrsim 1$) and the period-three motion at resonance are robust and were always observed. We speculate that further stable and also unstable states (solutions of the Navier--Stokes equations) may exist in phase space. A full understanding of the high-dimensional dynamics observed here poses a challenge to be tackled with the direct numerical computation of periodic solutions of the governing equations \citep{Kawahara2012}.

The amplitude of the analytic solutions of the Duffing oscillator \eqref{eq:duffing_amplitude} are plotted as lines in figure~\ref{fig:resonance_states}(\textit{a}). The measurements in (\textit{a}) follow closely the stable solutions (solid lines) with the important exception that the jump-down transition (and therefore the response maxima) occurs at substantially higher driving frequencies in the experiment. The jump-down transition obtained from the Duffing oscillator fit is at $\Omega \approx 0.78$ with an amplitude of $\hat X=9.44 \%$. In the sloshing experiments such high amplitudes are not reached, suggesting that enhanced dissipation prevents the development of such flow states.  

In undamped (inviscid) systems the phase-lag between excitation and response is zero. The phase-lag of the Duffing oscillator   \eqref{eq:duffing_phase} is plotted as lines in figure~\ref{fig:resonance_states}(\textit{b}), together with the experimentally measured values. Note, that only the sloshing amplitude was used to obtain the fitting parameters and not the phase difference. It can be seen, that the phase-lag of the low-amplitude sloshing with quasi-planar waves (marked as green triangles), as well as of the plateau state, are again in excellent agreement with the Duffing oscillator fit. However, the phase-lag of the larger amplitude sloshing states (peak state, wave-breaking, period-three) substantially deviates from the Duffing curve. This suggests that the damping is nonlinear for these complex flow states. 

The transition from the period-three motion --dominating close to the response maxima -- to the quasi-planar waves occurs at about $90^\circ$. Indeed, the phase-lag measurements shown in figure~\ref{fig:resonance_states}(\textit{b}) evidence that there are two distinct period-three states -- both exhibiting the jump-down transition at about $90^\circ$. 

We conclude that the sloshing amplitude (defined by the liquid's centre of mass) is neither suited to distinguish between flow states, nor does it act as an early indicator for the response maxima -- as the jump-down transition appears without warning. The phase-lag, on the other hand, allows a better distinction of the flow states and of the deviations from Duffing dynamics as the jump-down transition is approached. It seems that the jump-down transition occurs at a phase-lag of $90^\circ$, which we further test at different driving amplitudes in the next section.

\section{Response curves of sloshing for varying driving amplitudes}\label{sec:effectA}

\subsection{Comparison of the experiments to the response curves of a Duffing oscillator}

\begin{figure}
  \centering
  \includegraphics[width=13.5cm]{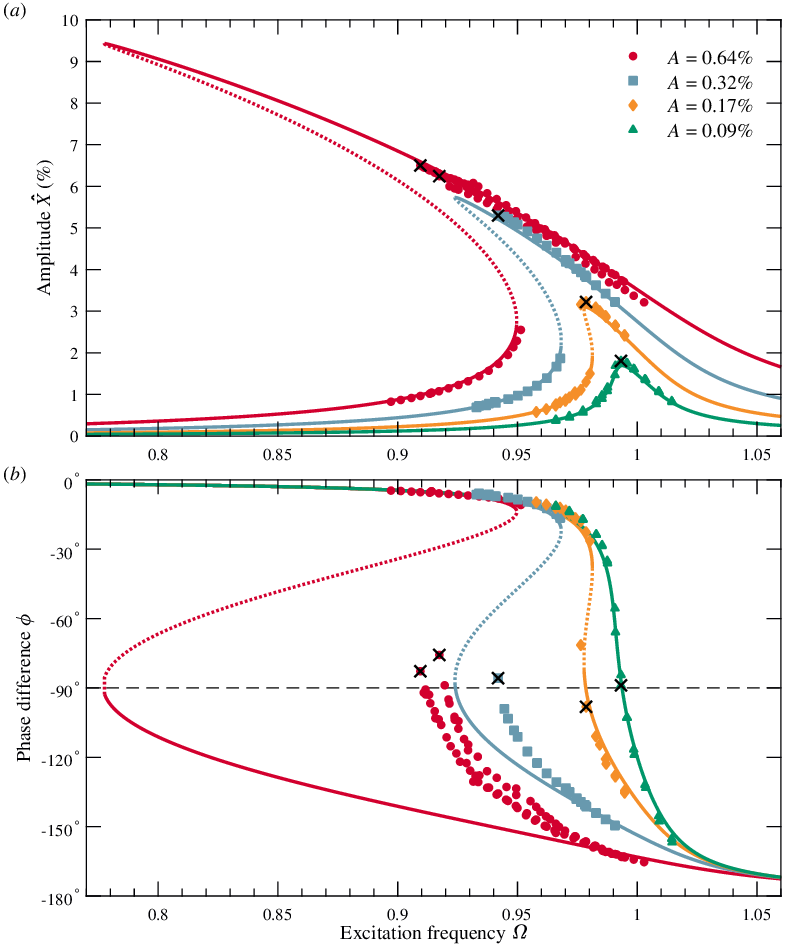}\\
  \caption{
    Quantitative comparison between sloshing experiments (symbols) and Duffing oscillator (lines) with a single free fitting parameter, $\varepsilon=-59.2$ in \eqref{eq:duffing_dimless}. (\textit{a}) Sloshing amplitude $\hat X$ exhibiting the characteristic bending of the resonance curve of a softening spring and the development of hysteresis for increasing excitation amplitude $A$.  Solid (dotted) lines denote stable (unstable) solutions of the Duffing equation. The measurements (symbols) were obtained by quasi-statically changing the excitation frequency $\Omega$, whilst keeping the excitation amplitude $A$ fixed. The experimentally determined maximum sloshing amplitudes are indicated by crosses ($\times$).   At $A=0.64\%$ the upper branch is composed of at least three branches of competing states (see figure~\ref{fig:resonance_states}). (\textit{b}) Phase-lag between sloshing and excitation. The transition from high to low amplitude sloshing occurs at a phase difference of $\phi \approx -90^{\circ}$.
  }
  \label{fig:resonance_driving}
\end{figure}

In this section, we quantitatively compare our experimentally obtained response curves for three additional driving amplitudes, $A=0.32\%$, $0.17\%$ and $0.09\%$ to the Duffing oscillator. The only parameter of the Duffing equation in \eqref{eq:duffing_dimless} that could neither be obtained from potential theory nor measured directly is the nonlinearity constant $\varepsilon=-59.2$, which was freely fitted. The response curves are shown in figure~\ref{fig:resonance_driving}, where experimental measurements are represented by symbols and the solid (dotted) lines denote stable (unstable) solutions of the Duffing equation.

At the lowest driving amplitude investigated, quasi-planar waves are found across the whole frequency range. The system behaves here nearly like a harmonic oscillator, but with the maximum sloshing amplitude slightly below the linear prediction ($\Omega=1$), and with the response curve marginally asymmetric (see figure~\ref{fig:resonance_driving}(\textit{a})). Both features are perfectly captured by the Duffing solution. For increasing driving amplitude hysteresis emerges, and the onset of hysteresis is in agreement with the analytical nonlinearity threshold for the Duffing oscillator \citep[$A_c\approx 0.13\%$, see][]{Brennan2008}. In the lower branch, the waves remain quasi-planar (as is the case for all $A$), whereas in the upper branch the wave peaks becomes higher and slightly more pointed. All our analyses suggest that the flow behaves here like an ideal Duffing oscillator, which is also reflected in the high quality of the fit (with discrepancies at the level of measurement uncertainty). 

At $A=0.32\%$ the agreement with the fit is still excellent, but in the experiments the jump-down transition occurs much earlier than in the Duffing curve. This behaviour becomes more pronounced at $A=0.64\%$, whereas the agreement in the jump-up transition point ($\Omega \approx 0.95$) remains excellent. In the Duffing oscillator, the jump-up transition depends on the nonlinearity only, whereas the jump-down transition depends also on the damping \citep{Brennan2008}. The premature jump-down transition in our experiments suggests a nonlinear increase of the damping, which is in line with the emergence of highly dissipative waves (with liquid steeply rising at the wall or wave-breaking, see figure~\ref{fig:states}(\textit{c}, \textit{d})) close to resonance.

As shown earlier, a more useful characterization of the deviations from Duffing dynamics is provided the phase-lag between excitation and response (shown in figure~\ref{fig:resonance_driving}(\textit{b})). For all driving amplitudes, we find that the phase-lag deviates from the Duffing prediction well before the resonance is reached.  At the resonance (and thus at the jump-down transition) the phase of the sloshing always lags the driving by $90^\circ$. Interestingly, in Duffing oscillators this jump-down transition at the response maxima occurs exactly at $90^\circ$ and it seems that this feature persists in our experiments despite the complex flow dynamics and the increased damping.

\subsection{Comparison of the experiments to the response curves of the multimodal model} \label{sec:response_multi_com}

\begin{figure} 
	\centering
	\includegraphics[width=13.5cm]{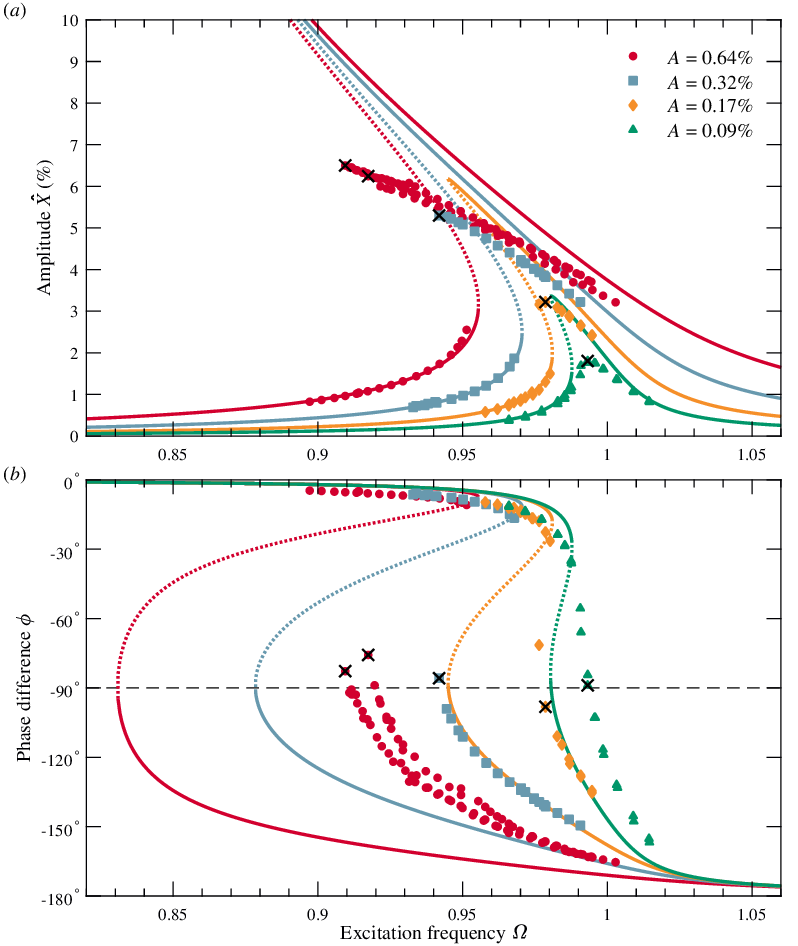}
	\caption{
	Comparison between the response curves from multimodal model (lines) and the experimentally measured values (symbols). (\textit{a}) Response of the center of mass amplitude  for different forcing amplitudes $A$. For the curves from the model the center of mass location is computed using \eqref{eq:multi:com} including the first three modes $N=3$. The amplitude $\hat X$ is selected as the maximum displacement over one oscillation period $t \in [0,T]$. (\textit{b}) Response curve of the phase-lag between sloshing oscillation and excitation from the multimodal model with \eqref{eq:multi:phase} and from the experiment.
	}
	\label{fig:multi_com}
\end{figure}

In this section, we quantitatively compare our experimentally obtained response curves to the analytical predictions of the multimodal model. We recall that the damping rate was determined from experimental measurements (see \S\ref{sec:damping}), as for the Duffing equation, whereas all other parameters are predicted by the theory and cannot be freely fitted. As shown earlier, the horizontal motion of the liquid's centre of mass $X$ provides an unambiguous determination of the sloshing amplitude and phase from experimental measurements, and is less sensitive to the exact wave shape than the surface elevation at a particular location. The horizontal oscillation of the liquid's center of mass $X$ can be computed from the multimodal model as 
\begin{equation} \label{eq:multi:com}
	X_N(t) = \frac{1}{w\,h} \int_{-w/2}^{w/2} \xi\left[ \sum_{n=1}^{N} \left(w \beta_n(t) f_n(\xi) \right) + h\right] \,d\xi \, /w.
\end{equation}
The response curve of the first three modes ($N=3$) is shown together with the experimental measurements in figure~\ref{fig:multi_com}.

At the lowest driving amplitude $A=0.09\%$ the model agrees well with the measurements sufficiently far from the resonance. However, in its vicinity no hysteresis is observed in the experiments and the response maxima reaches only about half the value of the model prediction. We stress that  at these parameters the flow dynamics is almost linear (with quasi-planar surface waves and frequency spectra similar to the one shown exemplarily in figure~\ref{fig:states}(\textit{i})). For increasing driving amplitudes $A$ the lower branch and its jump-up transitions are well described in the multimodal model. Here the wave shape is almost harmonic (as in figure~\ref{fig:multi_surface_snaps}(\textit{a})) leading to the excellent agreement of the response. At the upper branch, the sloshing amplitude obtained from the experiments is systematically below the model prediction (between $8\%$  and $30\%$) with growing deviations as the nonlinear response maxima are approached.

The damping rate is the only parameter not derived rigorously in the multimodal model. Modifying this value has little influence on the sloshing regimes, where the model predictions agree well with the experiments (e.g.\ at the lower branch solution and at the jump-up transitions). By contrast, the value of the damping ratio significantly affects the resonance amplitudes and the occurrence and the size of the hysteresis. If the damping ratio is decreased (e.g.\ $\gamma = 5.57 \cdot 10^{-3}$  from Keulegan's theory), the discrepancies with the experiment worsen. On the other hand, selecting larger values of the damping ratio (e.g.\ $\gamma = 15 \cdot 10^{-3}$), leads to better predictions of the hysteresis regions, especially at low amplitudes ($A=0.09\%, 0.16 \%$). The prediction of high-amplitude sloshing remains however poor independently of the value of $\gamma$. These observations indicate that the assumption of linear damping (applied only to the first mode) is at the root of causing the deviations from the experimental observations. This is further confirmed when comparing the phase difference $\phi$ between the driving and the sloshing response, shown in figure~\ref{fig:multi_com}(\textit{b}). Overall discrepancies are very large, even in regimes in which the sloshing amplitude $\hat X$ is reasonably predicted, and increase with the driving amplitude.

\section{Assessment of modelling approaches to sloshing}\label{sec:modelling}
In this section, we provide a brief summary and discussion of fundamentally different modelling approaches to sloshing with respect to our experimental observations. The Duffing equation and the multimodal model have already been assessed quantitatively in the preceding sections. Key to satisfactory comparisons between theory and experiment, is the direct measurement of the motion of the liquid's center of mass in the experiments, which was carried out in this work for the first time. The third and only recently developed approach (spectral submanifold theory) is substantially more general (and thus abstract) and has to our knowledge not been compared to experiments so far. We believe that our experiments can be equally stimulating to these three approaches, and to other modelling approaches. 
\subsection{Potential flow theory: Multimodal model}
The multimodal model developed by \cite{Faltinsen2009} is based on potential flow theory. It is analytical and does not have free fitting parameters. It is able to predict the full free-surface elevation and response curves for varying filling ratios, where for example the resonance shifts from softening to hardening. The multimodal analysis applied here is a single-dominant system with three degrees of freedom and has therefore a solid physical foundation. Intrinsically, dissipation and damping are not included because of the underlying potential-flow equations. Typically, a linear viscous damping is added to the equation for the first mode only. The corresponding damping coefficient can either be experimentally determined or analytically estimated following \citet{Keulegan1959}. In our study, the experimentally determined coefficient leads to a better agreement of the multimodal response curves with the experimentally measured ones. 
In general, the model only gives accurate amplitude predictions on the lower branch or far the resonance. The sloshing amplitude  at the resonance peak and the size of the hysteresis band are strongly overestimated even for small driving. The predicted phase difference between driving and response substantially deviates from the experimental measurements. 

Our measurements indicate that the assumption of linear damping applied to the first mode only is a plausible cause of the observed deviations. As the driving increases, the amplitude of the second and third modes increases and the damping associated to them likely become non-negligible. We however stress that the multimodal model provides the best purely analytical, nonlinear prediction of the sloshing behaviour. It is, for example, able to describe the competition and interaction between modes in shallow systems or in other tank geometries (e.g. of cylindrical form). 
Our rectangular tank geometry and large filling height, on the other hand, were selected to generate a sloshing dynamics with a well-distinguished response peak (i.e. free of mode competition) and a quasi-two dimensional flow. This simple setting paves the way for modelling the dynamics with the two following approaches.
\subsection{Mechanical mass-spring model: Duffing oscillator}
Simple mass-spring models have long been used in aerospace engineering to describe sloshing \citep{Dodge2000}. The (linear) harmonic oscillator emerges from a mechanical analogy, which enables the analytical calculation of the equivalent forcing amplitude and the equivalent spring constant. The Duffing equation extends the harmonic oscillator by including a nonlinear (cubic) spring deformation and therefore lacks an a-priori justification for modelling sloshing. A surprisingly good agreement between response curves of the amplitudes is achieved with a single free fitting parameter (e.g.  the nonlinearity constant $\epsilon$). Measuring a single response curve in the experiments is sufficient to obtain $\epsilon$, and therefore also the response curves at other driving amplitudes (including the correct threshold value for the resonance to become hysteretic). However, at large driving amplitude the height of the resonance peak and the size of the hysteresis are overestimated. The predicted phase-lag between driving and response agrees well at low sloshing amplitudes and deviates progressively for increasing amplitude. Although the agreement of the experiments with the Duffing oscillator is overall excellent, it is likely to fail for a further increasing driving amplitude, low filling height or other tank geometries. These would lead to highly complex effects and internal resonances that cannot be described by the Duffing equation anymore. Our study and the discovery of the Duffing dynamics might open avenues for modelling approaches that have the potential to capture and predict such effects.  One of the most promising approaches is described in the following.
\subsection{Dynamical systems approach: Spectral submanifold theory}
Sloshing is described by an evolutionary partial differential equation (the Navier--Stokes equation), i.e. formally by an infinite-dimensional dynamical system \citep{Ockendon1973}. The main idea of the spectral submanifold theory is that the relevant dynamics may be lower dimensional and limited to a spectral submanifold. The excellent agreement of our experiments (at low to moderate driving amplitudes) with the forced-damped Duffing equation can only be explained by such a behaviour. It seems that the sloshing dynamics of our system is confined to a low-dimensional attractor on which the leading-order dynamics is just the Duffing equation. The recently introduced concept of time-periodic, spectral submanifolds guarantees in these types of oscillatory problems precisely a two-dimensional, time-periodic, attracting invariant manifold \citep{Breunung2018}. The increasing deviations from the Duffing equation at larger driving amplitudes indicate that higher-order polynomial models are required to approximate the dynamics on the reduced two-dimensional, time-periodic spectral submanifold. In principle these polynomials could be fitted to the experimental data, but this approach is out of the scope of this work. Note that in our experiment internal resonances seem to be absent, though an exception could be the reported period-three motion. These are likely to occur in other tank geometries or at lower filling levels and would lead to a dynamics that cannot be described anymore in a two-dimensional spectral submanifold. In that case, higher-dimensional spectral submanifolds (with coupled Duffing-type equations) might provide again accurate predictions.

\section{Conclusions}\label{sec:conclusion}

Accurately describing and modelling the sloshing motion of liquids is an important topic in aerospace engineering and an outstanding scientific challenge. In previous works, several different approaches have been employed to investigate and even control sloshing. In this paper, we provide a detailed analysis of the sloshing dynamics in a deep rectangular tank with the overarching goal to bring together the theoretical fundamentals of nonlinear dynamical systems with standard methods used to investigate sloshing in tanks of e.g.\ space vehicles.

For fundamental studies, as well as for applications, the observation of the nonlinear resonance maxima in laboratory experiments is the key point. We stress the importance of measuring the response for several forcing amplitudes, because as pointed out by  \citet{Breunung2018} `a single response curve for a given forcing is meaningless for different forcing amplitudes' given the `essential nonlinear relationship between forcing and response amplitude of nonlinear systems'. We  performed measurements for four driving amplitudes. The dynamics ranges from an almost ideal Duffing oscillator with a softening spring, up to complex dynamics with competing modes, wave-breaking and period-three motion. We showed that the true nonlinear resonance maxima can only be detected if the dependence on the initial conditions is carefully considered, as done here and in the past (see e.g. \cite{Abramson1966a}) with a sweep-up, sweep-down procedure. By contrast, in many recent investigations of spacecraft tanks, the response curves are obtained by starting all experiments from rest, which may severely underestimate the maximum possible response of the system. Noteworthy, Duffing already highlighted the relevance of such `hidden' nonlinear resonances for engineering applications by stating that `when one wants to get out of a danger, it is good to know which way one has to go.' \citep[][translated from German by \citet{Kovacic2011}]{Duffing1918}. This must be considered in active feedback controls, as currently in development for rockets \citep{Konopka2019}, where the excitation of an unknown resonance state with the active control could be catastrophic.

By obtaining the motion of the liquid's centre of mass from our flow visualisation and using it to characterise the sloshing amplitude we were able to unambiguously measure the damping rate and the phase-lag between driving and response. Compared to classic single point-measurements of the surface height, our procedure drastically reduces the scatter of the experimental data (e.g. stemming from varying flow states). The motion of the liquid's centre of mass can equally be  computed from the multimodal model of  \cite{Faltinsen2009} and  is ideally suited to compare to nonlinear mass-spring models, such as the Duffing oscillator. In this paper, we presented  quantitative comparisons of unprecedented quality between experiment and models. Deviations between the models and the experiment significantly increase with the sloshing amplitude. Arguably, to model the surface waves at larger sloshing amplitudes correctly substantially more modes would be required in the multimodal analysis. However, the comparison with the Duffing oscillator indicates that damping is probably more important. In the multimodal model the damping is only included in the first mode and our data suggest that this might be the main reason causing the deviations.

We hope that our measurements stimulate the further development of models. However, the modelling of wave-breaking and run up of liquid on the tank walls, as often occurring in engineering application, still poses a great challenge and hinders a reliable prediction of the nonlinear response maxima. Our experiments provide here a new perspective and show that neither the exact surface shape, nor the frequency spectrum are the key indicators for the response maxima. Instead, we find the phase-lag between the sloshing and the driving is an excellent predictor of the nonlinear resonance: independently of the flow state (and thus of the degree of dissipation) the response is always delayed by $90^\circ$ at resonance. This remarkable behaviour is perhaps our most important observation reported here and was in fact theoretically predicted to occur in all periodically driven mechanical systems having `any damping that is a polynomial function of the velocities and positions' \citep{Breunung2018}. Very recently a rigorous mathematical argument that covers both primary and subharmonic resonances was obtained also for systems, without the assumptions of synchronous motion and linear damping by \cite{Cenedese2020}. Originally, this so-called $90^\circ$-phase-lag criterion was developed for the modelling of the primary resonance of structure vibrations \citep{Peeters2011}, but it is quickly gaining attention, because it is highly relevant for the determination of the backbone curve (which connects the resonance maxima for varying driving frequency and amplitude). For experimentalists it serves as an in situ-indicator, if the response maxima is reached. It would be interesting to test whether this behaviour is also found for other tank geometries (e.g.\ cylindrical) as employed in aerospace engineering.

%% additional chapters
\section*{Acknowledgements}
We thank Michael Dreyer and his team for support and Marc Avila for discussions regarding sloshing models. B. B. acknowledges funding for a `Bridge Scholarship' and K. A. for an `Independent Project for Postdocs' from the Central Research Development Fund of the University of Bremen. Funding was also provided from the Deutsche Forschungsgemeinschaft for the research unit FOR 2688 grant no. 417990176. 
%B. B. performed the experiments and analysis of the data. K. A. conceptualized and supervised the research. Both authors wrote the manuscript.

\section*{Declaration of interests}
The  authors  report  no  conflict  of  interest.

\section*{Supplementary movies}
Supplementary movies are available at [URL]

\appendix
\section{Comparison of the predicted and measured response curves based on a single-point surface elevation}

\begin{figure} 
	\centering
	\includegraphics[width=13.5cm]{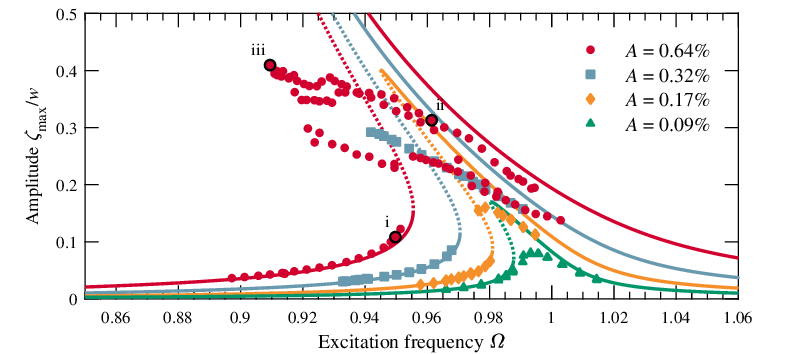}
	\caption{Comparison between the response curves from multimodal model (lines) and the experimentally measured values (symbols) based on the maximum dimensionless surface elevation at the tank wall,  $\zeta_{max}$. In the multimodal model, this quantity is calculated evaluating the first three modes ($N=3$) of \eqref{eq:multi:surf}. In the experiments, the value is obtained from the surface identification of the image processing. Surface waves of the multimodal model and the experiment are shown in figure~\ref{fig:multi_surface_snaps} at the parameters denoted with i, ii and iii. 
	}
	\label{fig:multi_surface_resonance}
\end{figure}

In \S\ref{sec:response_multi_com}, we compared the response curves of the multimodal model with those measured in our experiments. The comparison was based on the amplitude of the liquid's centre of mass. In this section, we apply the more commonly used method of evaluating the sloshing amplitude as a point-measurement of the surface elevation at the tank wall. As discussed in \S\ref{sec:com_vs_surface}, the surface displacement between wave crests and troughs can vastly differ, which leads to difficulties in unambiguously determining an amplitude from the experiments. The same applies to the multimodal model function of the surface displacement \eqref{eq:multi:surf}, where the addition of higher modes causes the difference between crest and troughs (see \S\ref{sec:multi_waves}). We here define sloshing amplitude as the maximum surface displacement $\zeta_{max}$ at the outer wall over one oscillation period ($t \in [0,T]$). The resulting response curves are shown in figure~\ref{fig:multi_surface_resonance}. 

At low sloshing amplitudes, the agreement between the multimodal model and the experiments is satisfactory and similar to the analysis based on the liquid's centre of mass (displayed in figure~\ref{fig:multi_com}). Moreover, the nonlinear response maxima occur at the same excitation frequency in both analyses. By contrast, at the largest driving amplitude $A=0.64\%$ the characterisation of the sloshing amplitude does have a significant impact on the scatter of the experimental data. Several flow states coexist at these parameters with very similar mass displacements, but largely differing surface elevations at the tank walls. The latter causes the scatter of the experimental data presented here and hinders a reliable assessment of the prediction of the multimodal model.

\bibliographystyle{jfm}
\bibliography{duffing}

\end{document}